\documentclass[prd,aps,showpacs,nofootinbib,eqsecnum,amsfonts,amsmath,
preprintnumbers,letterpaper,twocolumn]{revtex4-1}

\usepackage{graphics}
\usepackage{graphicx}
\usepackage{bm}
\usepackage{color}
\usepackage{amssymb}
\usepackage{psfrag}
\usepackage{times}
\usepackage{mathptmx}
\usepackage{hyperref}

\begin{document}

\newcommand{\be}{\begin{equation}}
\newcommand{\ee}{\end{equation}}
\newcommand{\ber}{\begin{eqnarray}}
\newcommand{\eer}{\end{eqnarray}}
\newcommand{\bea}{\begin{eqnarray}}
\newcommand{\eea}{\end{eqnarray}}
\newcommand{\ie}{i.e.}
\newcommand{\dt}{{\rm d}t}
\newcommand{\df}{{\rm d}f}
\newcommand{\dN}{{\rm d}N}
\newcommand{\dtheta}{{\rm d}\theta}
\newcommand{\dphi}{{\rm d}\phi}
\newcommand{\rhat}{\hat{r}}
\newcommand{\iotahat}{\hat{\iota}}
\newcommand{\phihat}{\hat{\phi}}
\newcommand{\hc}{{\sf h}}
\newcommand{\etal}{\textit{et al.}}
\newcommand{\balpha}{{\bm \alpha}}
\newcommand{\bbeta}{{\bm \psi}}
\newcommand{\fcut}{f_{3}}
\newcommand{\fmerg}{f_{1}}
\newcommand{\fring}{f_{2}}
\newcommand{\flow}{f_{\rm low}}
\newcommand{\rmi}{{\rm i}}
\newcommand{\matA}{{\sf A}}
\newcommand{\vecB}{{\bf B}}
\newcommand{\Mchirp}{M_{\rm c}}
\newcommand{\veclambda}{{\bm \lambda}}
\newcommand{\vectheta}{{\bm \theta}}
\def\rd{{\textrm{\mbox{\tiny{RD}}}}}
\def\qnr{{\textrm{\mbox{\tiny{QNR}}}}}
\newcommand{\A}{\mathcal{A}}
\newcommand{\NS}{\mathrm{NS}}
\newcommand{\lambdag}{\lambda_g}

\newcommand{\red}[1]{\textcolor{red}{{#1}}}
\newcommand{\ac}[1]{{#1}}

\newcommand{\LIGOCaltech}{LIGO Laboratory, California Institute of Technology, 
Pasadena, CA 91125, USA}
\newcommand{\TAPIR}{Theoretical Astrophysics, California Institute of
Technology, Pasadena, CA 91125, USA}

\title{Constraining the mass of the graviton using coalescing black-hole binaries}

\author{D.~Keppel}
\email{drew.keppel@ligo.org}
\author{P.~Ajith}
\email{ajith@caltech.edu}
\affiliation{\LIGOCaltech}
\affiliation{\TAPIR}


\bigskip

\begin{abstract}
We study how well the mass of the graviton can be constrained from
gravitational-wave (GW) observations of coalescing binary black holes. Whereas
the previous investigations employed post-Newtonian (PN) templates describing
only the inspiral part of the signal, the recent progress in analytical and
numerical relativity has provided analytical waveform templates coherently
describing the inspiral-merger-ringdown (IMR) signals. We show that a search
for binary black holes employing IMR templates will be able to constrain the
mass of the graviton much more accurately ($\sim$ an order of magnitude) than a
search employing PN templates. The best expected bound from GW observatories
($\lambdag > 7.8\times10^{13}$ km from Adv.~LIGO, $\lambdag > 7.1\times10^{14}$
km from Einstein Telescope, and $\lambdag > 5.9\times10^{17}$ km from LISA) are
several orders-of-magnitude better than the best available
\emph{model-independent} bound ($\lambda_g > 2.8 \times 10^{12}$ km, from Solar
system tests). 
\end{abstract}


\maketitle

\begin{figure*}[tbh]
\centering
\includegraphics[width=3.5in]{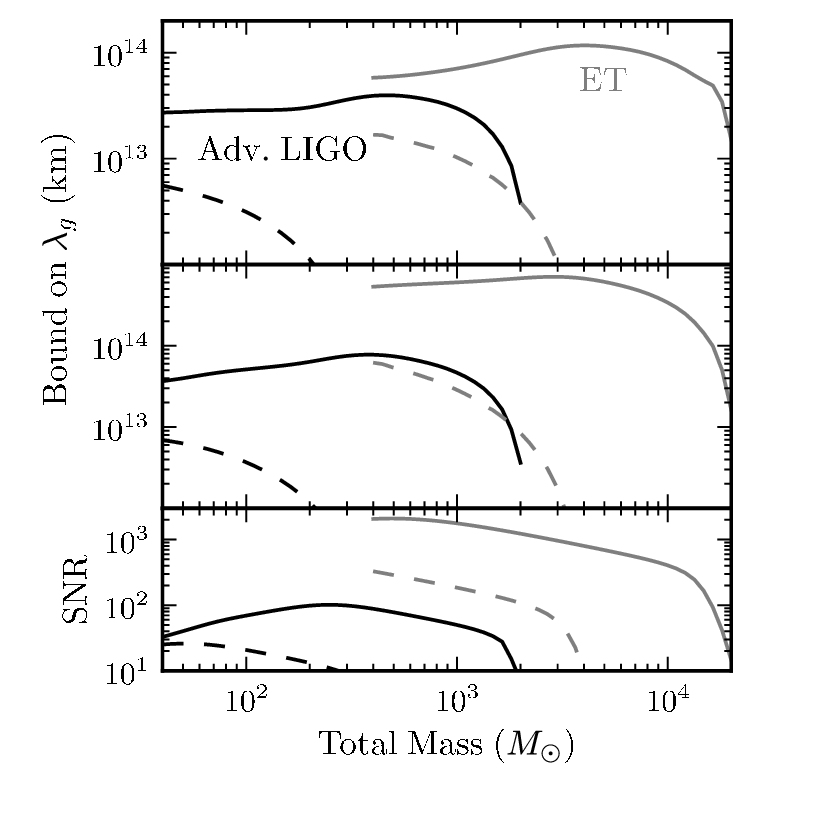}
\includegraphics[width=3.5in]{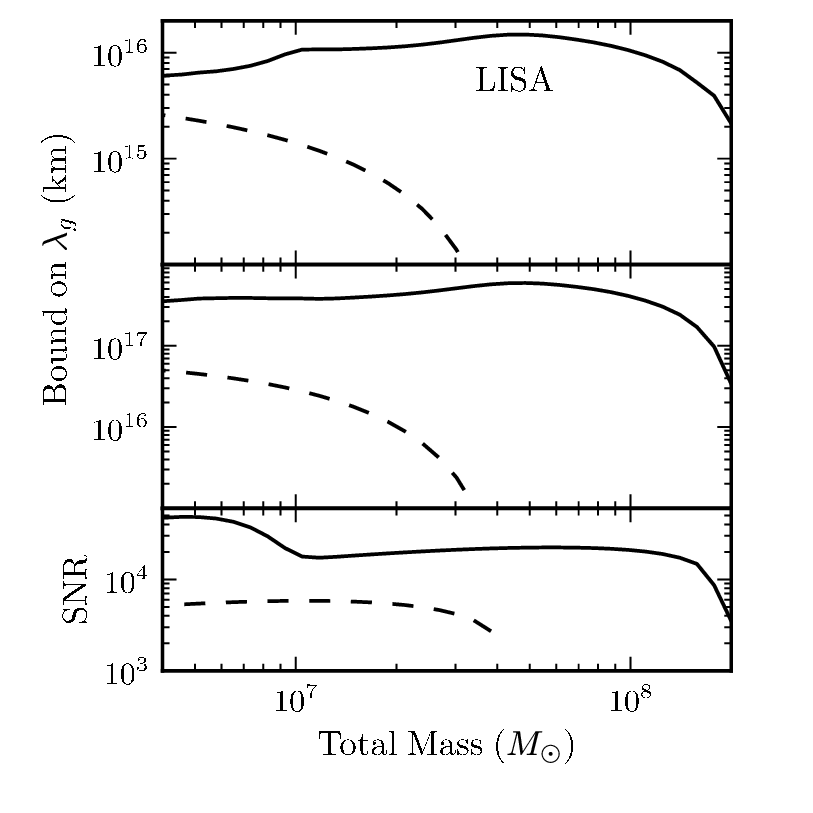}
\caption{\emph{Left.} Top panels show the lower bound on the Compton wavelength
$\lambdag$ of the graviton that can be placed from observations of equal-mass
binaries located at distances such that they produce optimal SNRs of 10 in the
Adv.~LIGO (black traces) and ET (grey traces) detectors using their smallest
low-frequency cutoffs (10 Hz and 1 Hz, respectively). Middle panels show the
same bounds from binaries located at 1 Gpc, and the bottom panels show the
optimal SNR produced by these binaries. Horizontal axes report the total mass
of the binary.  Solid and dashed lines correspond to IMR and restricted 3.5PN
waveforms, respectively.  \emph{Right.} Same plots for the case of binaries
located at 3 Gpc detected in the LISA detector.}
\centering
\label{fig:summaryplot}
\end{figure*}

\section{Introduction and Summary}\label{sec:Intro}

General relativity (GR) has proven to be very successful in predicting and
explaining a variety of gravitational phenomena. Over the last century, basic
ingredients of GR have been tested in many different ways and in many different
settings. In GR, gravitational interactions propagate with the speed of light,
which means that the hypothetical quantum particle of gravity, the
``graviton'', has no rest mass. A non-zero graviton mass $m_g$ would produce
several interesting effects: for e.g., it will cause the gravitational
potential to take the Yukawa form, effectively cutting off gravitational
interactions at distances greater than the Compton wavelength $\lambdag \equiv
h/m_gc$ of the graviton. Indeed, the absence of such effects in the solar
system has provided a lower bound on $\lambdag$ (hence an upper bound on
$m_g$). But a graviton mass with corresponding Compton wavelength much larger
than the size of the solar system has not yet been ruled out in a
model-independent way\footnote{It has been argued that ``dark energy''
\emph{might} be explained by a graviton-mass-like effect, with associated
Compton wavelength comparable to the radius of the visible universe.
See~\cite{Goldhaber:2008xy} for a review.}. Also, if the graviton has non-zero
mass, the gravitational waves (GWs) will have extra degrees of freedom (such as
longitudinal modes), and will travel with a frequency-dependent speed,
different from the speed of light.  

Whereas the current experimental bounds on theories of gravity are from the
weak-field limit, GW observations provide excellent test beds in the dynamical,
strong-field regime  (see \cite{lrr-Will:2006} and \cite{lrr-2009-2} for recent
reviews). Indeed, these are exciting times for the word-wide GW community. The
Initial LIGO~\cite{Abbott:2007kv} detectors have completed their fifth science
run, at design sensitivity. The Virgo~\cite{VirgoStatus-GWDAW2008} and GEO\,600
detectors ran concurrently with LIGO for part of that run. Although a direct
detection of GWs is yet to be made, a number of interesting astrophysical upper
limits are constructed based on this data (see e.g.,~\cite{Abbott:2008fx,
Collaboration:2009ws}). After the ongoing sixth science run at a further
improved sensitivity, the LIGO-Virgo detectors will undergo a commissioning
break with the target of achieving much improved sensitivities with Advanced
LIGO and Advanced Virgo.  Implementing a number of advanced technologies
including the use of squeezed light, the GEO\,600 is already undergoing
commissioning work for the high-frequency detector configuration
GEO-HF~\cite{Willke:2006uw}. Also, the design study for a third generation
detector, called the Einstein Telescope (ET) is ongoing in Europe. With the
ever closer approach of the era in which GW observations become routine, it is
interesting to see and update what information could be gathered about the
Universe from these observations. 

In this paper we investigate the possible bounds that can be put on theories
with an effective mass in the propagation of GWs (massive-graviton
theories\footnote{ It has to be mentioned that so far there is no complete
theory of gravity in which the graviton can have a mass. Constructing such a
theory in a ghost-free manner has proven to be nontrivial; see,
e.g.,~\cite{Yunes:2009ke} and references therein.}, in short) from GW
observations of coalescing binary black holes (BBHs). In massive-graviton
theories, the speed of propagation of GWs depends on the wavelength. If the GW
emission is accompanied by electromagnetic (EM) emission (such as the case of a
core-collapse supernova or a white-dwarf binary), a bound on the mass of the
graviton can be placed by comparing the time of arrival of the GW signal with
the EM signal~\cite{Will98}, or, by correlating the EM and GW
signals~\cite{Larson:1999kg,Cutler:2002ef,Kocsis:2008}.  But, since many of
these sources are not very well understood in terms of their emission
mechanisms and the delay between EM and GW emissions, this could introduce
significant uncertainties in such measurements.  

Coalescing BBHs provide us with the potential to constrain the mass of the
graviton without relying on the presence of an EM counterpart.  In the case of
BBHs, since the frequency of gravitational radiation sweeps from lower to
higher frequencies, the speed of the emitted gravitons will vary, from lower
speeds to higher speeds. This will cause a distortion in the observed phasing
of the GWs. Since BBHs can be accurately modelled using analytical/numerical
solutions of Einstein's equations, any deviation from GR can be parametrized
and measured (see, e.g.,~\cite{Yunes:2009ke} for a recent discussion on such a
general framework). A framework for testing this possibility by measuring the
distortion of GWs was originally developed by Will~\cite{Will98}.  Will found a
dispersive effect that appeared as an additional term in the post-Newtonian
(PN) expansion of the GW phase, and showed that a bound on the mass of the
graviton can be placed from the GW observations by applying appropriate matched
filters. 

Will's original work was performed using \emph{restricted} PN waveforms
describing the \emph{inspiral} stage of non-spinning coalescing compact
binaries, the phase of which was expanded to 1.5PN order. Recent work has
elaborated on this by incorporating more accurate detector models, and by
including more physical effects such as effects rising from the spin angular
momentum of the compact objects, from the eccentricity of the orbit, and from
the inclusion of higher harmonics rising from the contribution of the higher
multipoles \cite{Will:2004xi,BBW2004, Jones:2004uy,Berti:2005qd,
Stavridis:2009mb, YagiTanaka09, ArunWill09}.

Since the PN formalism has enabled us to compute accurate waveforms from the
\emph{inspiral} stage of the coalescence, these analyses have focused on the
information gained from the observation of the inspiral stage. The last few
years have witnessed a revolution in the numerical simulations of compact
binaries. In particular, numerical relativity was able to obtain exact
solutions for  the ``binary-black-hole problem'' \cite{Pretorius1, Baker05a,
Campanelli:2005dd}.  Concomitant with this great leap has been significant
progress in analytical relativity in the computation of high order PN terms and
the inclusion of various effects arising from spins, higher harmonics etc.
Combining the analytical and numerical results, different ways of constructing
inspiral-merger-ring-down (IMR) waveforms have been proposed
\cite{Ajith:2007kx,Buonanno:2007pf,Damour:2007yf}.  It has been widely
recognized that these IMR waveforms will significantly improve the sensitivity
and distance reach of the searches for BBHs (see, e.g., \cite{FlanHugh98,
Buonanno:2006ui, Ajith:2007kx}) as well as the accuracy of the parameter
estimation (see, e.g., \cite{Ajith:2009fz, Aylott:2009ya,
AylottVeitchVecchio:2009}).

In this paper, we estimate the bounds that can be placed on the mass of
graviton from the GW observations of BBHs using IMR templates. This is
motivated by the previous observations (see e.g.~\cite{Ajith:2009fz}) that the
IMR waveforms will significantly improve the accuracy of the parameter
estimation by breaking the degeneracies between the different parameters
describing the signal, including the parameter describing the mass of the
graviton.

Due to the intrinsic randomness of the noise in the GW data, the estimated
parameters of the binary (including the one parameter describing the mass of
the graviton) will fluctuate around their mean values.   In the limit of high
signal-to-noise ratios (SNRs), the spread of the distribution of the observed
parameters --- the accuracy of the parameter estimation --- is quantified by
the inverse of the \emph{Fisher information matrix}~\cite{Cramer46,Rao45}.  We
employ the Fisher matrix formalism to estimate the expected bounds on the mass
of the graviton using the \emph{non-spinning} limit of the IMR waveform model
proposed by Ref.~\cite{Ajith:2009bn}. This is a frequency-domain waveform
family describing the leading harmonic of the IMR waveforms from BBHs. 

Indeed, in the absence of a complete massive-graviton theory, we consider only
the propagation effects of a massive graviton, and assume that the wave
generation is correctly given by GR, apart from negligible corrections.  This
will introduce a systematic error on the possible constraints, which depends on
the specific massive graviton theory\footnote{If stable black holes exist in
such theory, then we might expect that the major effect comes from the
propagation, since the length scales involved in the propagation are typically
much larger than the length scales involved in the wave generation. However, it
is difficult to address this properly in the absence of a complete theory of
massive graviton.}. Since we are neglecting spins and higher harmonics in the
(general relativistic) signal model, this introduces another set of systematic
errors; but these can be circumvented by including these effects in the signal
model. We remind the reader that this paper only tries to quantify the
noise-limited statistical errors in the observations.

The main findings of the paper are summarized below
(Section~\ref{sec:summary}).  The following sections present the details of the
analysis. Section~\ref{sec:dispcbc} briefly reviews the effect of massive
graviton on the dispersion of GWs, and summarizes the existing bounds on the
graviton mass. In Section~\ref{sec:gwdispmgbounds}, we compute the expected
upper bounds that can be placed on the mass of the graviton using the
observations of IMR signals. In that section, we review the signal and detector
models used, provide the details of the computation and present a discussion of
the results and the limitations of this work. 

\subsection{Summary of results}
\label{sec:summary}

An executive summary of results is presented in Fig.~\ref{fig:summaryplot} for
the case of ground-based detectors Adv.~LIGO and ET as well as the space-borne
detector LISA.  For ground-based detectors, the binary is assumed to be located
optimally oriented at 1 Gpc, and for LISA, the binary is located at 3 Gpc. For
the case of Adv.~LIGO (with low-frequency cutoff, $\flow$ = 10 Hz), the best
bound ($\lambdag > 7.8\times10^{13} \mathrm{km} \simeq 2.5 \mathrm{pc}$; $m_g <
1.6\times10^{-23}$~eV) using IMR templates is obtained from the observation of
binaries with total mass $M \simeq 360 M_\odot$. This is significantly better
than the \emph{best} bound obtained using restricted 3.5PN templates ($\lambdag
> 7.9\times10^{12}$ km for $M\simeq 18 M_\odot$).  For ET (with $\flow$ = 1
Hz), the best bound using IMR templates ($\lambdag > 7.1\times10^{14}
\mathrm{km} \simeq 23 \mathrm{pc}$; $1.7\times10^{-24}$~eV) is obtained from
binaries with $M\simeq 3000 M_\odot$, while the best bound for PN templates
($\lambdag > 1\times10^{14}$ km) is obtained for binaries with $M\simeq 65
M_\odot$. For LISA observation of supermassive black-hole (BH) binaries, the
best bound ($\lambdag > 5.9\times10^{17} \mathrm{km} \simeq 19 \mathrm{kpc}$;
$2.1\times10^{-27}$~eV) using IMR templates is obtained from binaries with $M
\simeq 4.8\times10^{7}M_\odot$, while the best bound using PN templates
($\lambdag > 6.3\times10^{16}$ km) is obtained from binaries with $M\simeq
1.9\times10^{6} M_\odot$. In summary, the \emph{best} bounds using IMR
templates are roughly an order of magnitude better than the \emph{best} bounds
using restricted PN templates. The improvement is partly due to the higher SNR
and partly due to the extra information harnessed from the post-inspiral
stages. 

The best expected bound from ground-based observatories is over two
orders-of-magnitude better than the best available model-independent bound
($\lambdag > 2.8 \times 10^{12}$ km) given by monitoring the orbit of Mars (see
Sec.~\ref{sec:otherbounds}). Additionally, LISA will be able to improve on
those bounds by several orders of magnitude.  Most importantly, GW observations
will provide the first constraints from the highly dynamical, strong-field
regime of gravity. Table~\ref{tab:boundsGWsummary} summarizes the bounds that
can be placed with future GW observations employing different signal models /
analysis methods. 
 
\smallskip 

\noindent \emph{Effect of low frequency cutoff:---} The best bound that can be
placed with Adv.~LIGO noise spectrum with low-frequency cutoff, $\flow$ = 10 Hz
will be $\sim$ 25\% better than the same obtained with $\flow$ = 20 Hz. For ET,
the best bounds using a configuration with $\flow$ = 1 Hz is $\sim$ 70\% better
than the same obtained with $\flow$ = 10 Hz. We hope that this information can
contribute to weighing the scientific case of different configurations of
advanced detectors. 

\noindent \emph{Detectability of the signals by GR-based templates:---} If we
assume the constraints on the graviton mass given by solar system tests, the
``mismatch'' between the signal and GR-based templates is unacceptably high
(7--50\%). However, if the mass of the graviton is a factor of three smaller
than the solar system limit, then the mismatch is negligible. This provides a
rough estimate of the detectability of the deformed signals (due to the
graviton mass) using GR-based templates.

\begin{table}[tbh]
\begin{center}
\begin{tabular}{ccccc}
    \hline
    \hline
    Signal model / method &\vline& Adv.~LIGO       & ET & LISA  \\
    \hline
    \hline
    \parbox[t][0.13in]{1.2in}{\footnotesize 1.5PN inspiral~\cite{Will98}} &\vline& 0.6 (20) &  & 0.7 ($2\times10^7$)  \\
    \parbox[t][0.26in]{1.3in}{\footnotesize 1.5PN inspiral, more accurate detector model~\cite{Will:2004xi}} &\vline&  &  & 	0.5 ($10^7$)  \\
    \parbox[t][0.26in]{1.2in}{\footnotesize 3.5PN inspiral, no spin [this paper]} &\vline& 0.8 (18) & 	10 (65) & 	0.6 ($1.9 \times 10^6$)  \\
    \parbox[t][0.26in]{1.2in}{\footnotesize 3.5PN inspiral, higher harmonics, no spin~\cite{ArunWill09}} &\vline& 0.7 (60) & 	10 (400) & 	0.5 ($2\times10^6$)  \\
    \parbox[t][0.26in]{1.2in}{\footnotesize 3.5PN inspiral, 2PN spin, no precession~\cite{BBW2004}} &\vline&  & 	 & 	0.5 ($2\times10^6$)  \\
    \parbox[t][0.26in]{1.2in}{\footnotesize 2PN inspiral, spin precession~\cite{Stavridis:2009mb}} &\vline&  & 	 & 	0.7 ($2\times10^7$)  \\
    \parbox[t][0.26in]{1.3in}{\footnotesize 2PN inspiral, simple spin precession, eccentricity~\cite{YagiTanaka09}} &\vline&  & 	 & 	0.4 ($10^7$)  \\
    \parbox[t][0.13in]{1.2in}{\footnotesize IMR, no spin [this paper]} &\vline& 8 (360) & 	70 (3000) & 	6 (4.8 $\times 10^7$)  \\
    \parbox[t][0.39in]{1.2in}{\footnotesize Measuring phase of arrival of diff.~harmonics from eccentric binaries~\cite{Jones:2004uy}} &\vline&  & 	 & 	0.3 ($10^6$)  \\
    \parbox[t][0.39in]{1.2in}{\footnotesize Correlating GW and EM observations from white dwarf binaries~\cite{Cutler:2002ef}} &\vline&  & 	 & 	$10^{-3}$ (2.8)  \\
    \parbox[t][0.52in]{1.3in}{\footnotesize Correlating GW and EM observations from supermassive BH binaries~\cite{Kocsis:2008}} &\vline&  & 	 & 	$9 \times 10^{-2}$ ($10^{7}$)  \\
    \hline
    \hline
\end{tabular}
\caption{Expected bounds on $\lambdag$ (in units of $10^{13}$ km for the case
of Adv.~LIGO and ET, and $10^{17}$ km for LISA) to one significant digit using
future GW observations employing different signal models / analysis methods.
Total mass (in $M_\odot$) giving the bound are shown in brackets. It should be
noted that different investigations have used slightly different source- and
detector models, and hence, are not strictly comparable.}
\label{tab:boundsGWsummary}
\end{center}
\end{table}

\section{Dispersion of gravitational waves and the existing bounds on the mass of the graviton}
\label{sec:dispcbc}

\subsection{Dispersion of gravitational waves}

Will \cite{Will98} derived the dispersive effects of a massive graviton on the
propagation of GWs from coalescing compact binaries. If the gravitons are
massive, the GWs will travel with a speed $v_g$ different from the speed of
light\footnote{Throughout the rest of this paper, we use geometric units:
$G=c=h=1$.}, given by
\begin{equation}
v_g^2 = 1 - m_g^2/E_g^2 \equiv 1 - \left(\lambdag f\right)^{-2},
\label{eq:massivegravDispersion}
\end{equation}
where $m_g$ is the graviton rest mass, $E_g$ its rest energy, $\lambdag \equiv
m_g^{-1}$ its Compton wavelength, and $f$ is the frequency of the gravitational
radiation. Consider two gravitons with frequency $f_e$ and $f'_e$ emitted in
the time interval $\Delta t_e$ at the source. The dispersion relation given in
Eq.(\ref{eq:massivegravDispersion}) will change this time interval to 
\begin{equation}
\Delta t = (1+Z) \, \left[ \Delta t_e + \frac{D}{2 \lambdag^2} 
\left(\frac{1}{f_e^2}-\frac{1}{f_e^{\prime2}} \right)\right],
\end{equation}
when observed from a distance $D$. The distance parameter $D$ is related to the 
luminosity distance $D_L$ by
\begin{equation}
D = D_L \left[\frac{1 + (2+Z)(1+Z+\sqrt{1+Z})}{5(1+Z)^2} \right],
\end{equation}
where $Z$ is the cosmological redshift. Such a change in the arrival time will
lead to a distortion of the observed phasing of the GW signal at the detector,
as described by Eq.(\ref{eq:phaseDistortion}).  

\subsection{Existing bounds on the mass of the graviton}
\label{sec:otherbounds}

The most stringent available bound on the mass of the graviton (which does not
rely on specific massive-graviton theories) comes from the effect of a massive
graviton field on the static (Newtonian) gravitational potential, changing it
from $M/r$ to the Yukawa form $Mr^{-1} e^{-r/\lambdag}$. If this were the case,
Kepler's third law would be violated since the gravitational force would no
longer follow an inverse-square law. The absence of such an effect in the solar
system thus provides an upper limit on the graviton mass.  The most stringent
bound comes from the orbit of Mars which limits the mass of the graviton such
that $\lambda_g > 2.8 \times 10^{12}$ km.~\cite{Talmadge:1988}. 

Another, slightly less sensitive, bound is given by the binary pulsar
observations.  If the graviton had mass, the orbits of binary pulsars would
decay at a slightly faster rate than predicted by GR, due to additional energy
loss from the leading order massive graviton terms in the power radiated.
Combining the observations of PSR B1913+16 and PSR B1534+12,
Ref.~\cite{FinnSutton2002} obtained the $90\%$ confidence bound $\lambda_g >
1.6 \times10^{10}$ km. 

A number of model dependent, albeit more sensitive bounds have been constructed
by invoking assumptions such as specific massive-graviton theories and specific
distributions of dark matter. For a summary of such results, we refer the
reader the recent review~\cite{Goldhaber:2008xy}.

\section{Expected bounds on the mass of the graviton using inspiral-merger-ring-down waveforms}
\label{sec:gwdispmgbounds}

\begin{figure}[tbh]
\centering
\includegraphics[width=3.1in]{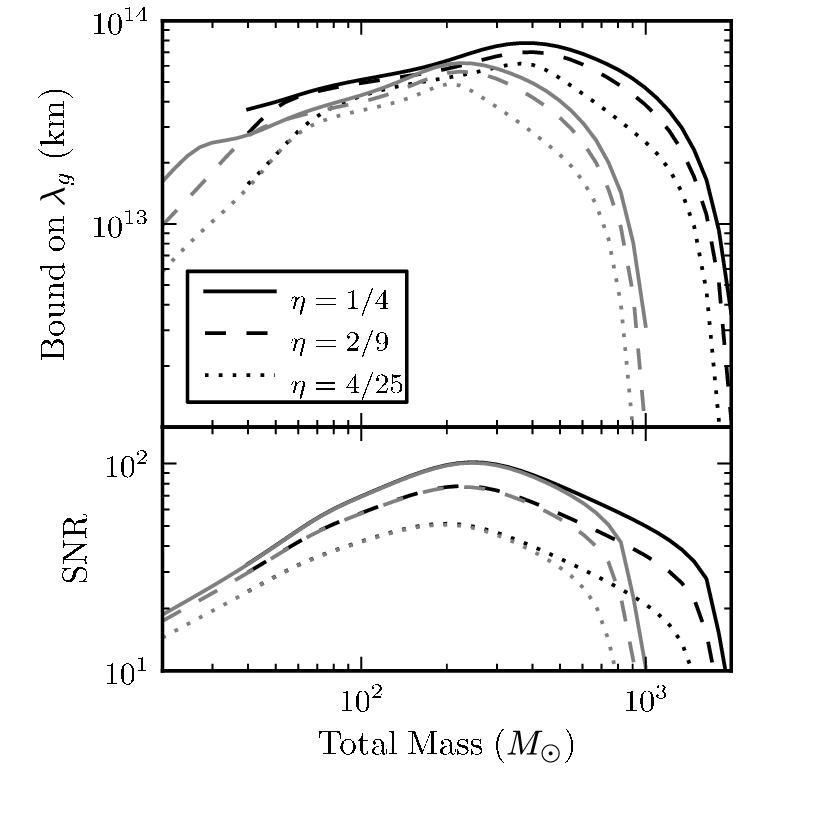}
\caption{Optimal SNR (lower panel) and the lower bound on the Compton
wavelength $\lambdag$ of graviton (upper panel) for IMR waveforms detected at
1~Gpc by the Adv.~LIGO detector.  Black and grey curves correspond to
low-frequency cutoffs 10 Hz and 20 Hz, respectively, while solid, dashed, and
dotted curves correspond to mass ratios $\eta$ = 1/4, 2/9 and 4/25.}
\label{fig:AdvLIGO_all}
\end{figure}

\subsection{Signal model}

The (general-relativistic) signal model we use is the \emph{non-spinning limit}
of IMR waveforms describing BBHs with non-precessing spins, proposed by
Ref.~\cite{Ajith:2009bn}. These frequency-domain waveforms can be written as
$\tilde{h}(f) \equiv A(f) \, e^{\rmi \Psi(f)}$, where the amplitude and phase
are given by
\ber
A(f) &\equiv& C \fmerg^{-7/6}
\left\{ \begin{array}{ll}
f'^{-7/6} \, (1+ \sum_{i=2}^{3} \alpha_i \, v^{i})   & \textrm{if $f < \fmerg$}\\
w_m \, f'^{-2/3} \, (1+ \sum_{i=1}^{2} \epsilon_i \, v^{i}) & \textrm{if $\fmerg \leq f < \fring$}\\
w_r \, {\cal L}(f,\fring,\sigma) & \textrm{if $\fring \leq f < \fcut$}\\
0 & \textrm{if $ f > \fcut$,}\\
\end{array} \right. \nonumber \\
\Psi(f) &\equiv& 2 \pi f t_0 + \varphi_0 + \frac{3}{128 \eta v^5} 
\big(1 + \sum_{k=2}^{7} v^{k} \, \psi_k \big).
\label{eq:phenWaveAmpAndPhase}
\eer
In the above expressions, $f'\equiv f/\fmerg$ and $v \equiv (\pi Mf)^{1/3}$,
where $M \equiv m_1+m_2$ is the total mass of the binary, and $\eta \equiv m_1
m_2/M^2$ is the symmetric mass ratio.  For non-spinning binaries, the PN
corrections to the Fourier domain amplitude of the ($\ell=2, m=\pm2$ mode) PN
inspiral waveform are $\alpha_2 = -323/224 + 451\,\eta/168$ and $\alpha_3 =
0$~\cite{Arun:2008kb}, while the phenomenological parameters describing the
merger amplitude take values $\epsilon_1 =  - 1.8897, \epsilon_2 = 1.6557$.
$C$ is a numerical constant depending on the location and orientation of the
binary, as well as the masses: $C=\frac{M^{5/6}}{D_L \pi^{2/3}} \sqrt{\frac{5
\eta}{24}}$ for optimally-oriented binaries, $D_L$ being the luminosity
distance.  The time of arrival of the signal at the detector is denoted by
$t_0$, the corresponding phase by $\varphi_0$, and $\mathcal{L}(f,f_2,\sigma)
\equiv \frac{1}{2 \pi} \frac{\sigma}{(f-f_2)^2 + \sigma^2/4}$ is a Lorentzian
function with width $\sigma$ centered around the frequency  $\fring$. 

For these waveforms, the amplitude of the ``inspiral'' part is described by the
amplitude of the PN inspiral waveform and the amplitude of the ringdown portion
is modelled as a Lorentzian, which agrees with the quasi-normal mode ringing of
a perturbed BH from BH perturbation theory. The merger amplitude is empirically
estimated from the numerical-relativity simulations.  The frequencies $f_1$ and
$f_2$ correspond to the transition points between the inspiral-merger and
merger-ringdown, and $f_3$ is a convenient cutoff frequency such that the
signal power above this frequency is negligible.  The normalization constants
$w_m$ and $w_r$ make ${A}(f)$ continuous across the transition frequencies
$\fmerg$ and $\fring$. 

The phase of the waveform is written as an expansion in terms of different
powers of the Fourier frequency $f$ (analogous to the phasing expression of the
PN inspiral waveform computed using the stationary-phase approximation).  But
the phenomenological phase parameters $\psi_k$ are tuned so that the analytical
templates have the best overlaps with the complete IMR waveforms.  Thus,
$\psi_k$ are different from the corresponding coefficients describing PN
inspiral waveforms. This also restricts the validity of these waveforms to the
range of GW frequency $f \gtrsim 2\times 10^{-3}/M$. In this paper, we only
consider binaries with total mass $M > 2\times 10^{-3}/\flow$, where $\flow$ is
the low-frequency cutoff of the detector sensitivity. 

The phenomenological parameters $\psi_k$ and $\mu_k \equiv \{\fmerg, \fring,
\sigma, \fcut\}$ are written in terms of the physical parameters $M$ and $\eta$
of the binary as:
\ber
\psi_k & = & \psi_k^0 + x_k^{(10)}\,\eta + x_k^{(20)}\,\eta^2 + x_k^{(30)}\,\eta^3 \nonumber \\ 
\pi M \mu_k  & = & \mu_k^0  + y_k^{(10)}\,\eta + y_k^{(20)}\,\eta^2 + y_k^{(30)}\,\eta^3,
\label{eq:ampParams}
\eer
where the coefficients $x_k$ and $y_k$ are tabulated in
Table~\ref{tab:IMRWaveCoeffs}.

The effect of a massive graviton is that it will create a distortion in the 
observed phasing of GWs, which can be written as \cite{Will98} 
\begin{equation}
\Psi_{\rm eff}(f) = \Psi(f) - \beta f^{-1} + \phi_g + \tau_g f,
\label{eq:phaseDistortion}
\end{equation}
where $\Psi(f)$ is given by Eq.(\ref{eq:phenWaveAmpAndPhase}). The terms
involving $\tau_g$ and $\phi_g$ will only result in a redefinition of the
measured arrival time $t_0$ and the phase offset $\varphi_0$, and hence can not
be independently measured. But the term involving $\beta \equiv \pi D/
\lambdag^2(1+Z)$ will produce an observable effect. In general, $\beta$ will
have non-zero correlation with other parameters describing the waveform, which
will limit our ability to estimate $\beta$ (see
Sec.~\ref{sec:ErrBoundsFromFishMat}).

For this calculation, we have assumed that the wave generation is given
correctly by GR. At least for the inspiral portion of the signal, it is
reasonable to expect that corrections to GR will be of the of order
$(r/\lambda_g)^2$, where $r$ is the size of the binary system \cite{Will98}.
Assuming the existing bounds on the graviton mass ($\lambda_g > 10^{12}$ km),
it can be shown that, for the binaries we are interested in, the correction is
negligible. But, if the BHs in the true massive-graviton theory is
significantly different from the BHs in GR, this could affect the wave
generation in the merger-ringdown stages. Since we do not have a complete
massive graviton theory, we are unable to address this issue at the moment.     

\begin{table}[tbh]
    \begin{center}
        \begin{tabular}{ccccccccccccccc}
            \hline
            \hline
            &\vline&    $\psi^0_k$ 	& $x^{(10)}$ 	& $x^{(20)}$ 	& $x^{(30)}$ \\
            \hline
            $\psi_2 $ &\vline& $\frac{3715}{756}$& -920.9 	            & 6742 	                & -1.34$\times 10^{4}$\\
            $\psi_3 $ &\vline& $-16 \pi$ 	     & 1.702$\times 10^{4}$ & -1.214$\times 10^{5}$ & 2.386$\times 10^{5}$\\
            $\psi_4 $ &\vline& $\frac{15293365}{508032}$ & -1.254$\times 10^{5}$& 8.735$\times 10^{5}$ 	& -1.694$\times 10^{6}$\\
            $\psi_6 $ &\vline& 0	             & -8.898$\times 10^{5}$& 5.981$\times 10^{6}$ 	& -1.128$\times 10^{7}$\\
            $\psi_7 $ &\vline& 0	             & 8.696$\times 10^{5}$ & -5.838$\times 10^{6}$ & 1.089$\times 10^{7}$\\
            \hline
            &\vline&  $\mu^0_k$   & $y^{(10)}$	& $y^{(20)}$ 	& $y^{(30)}$ \\
            \hline
            $\fmerg$ &\vline& $1 - 4.455 + 3.521$ & 0.6437 	 & -0.05822 	 & -7.092\\
            $\fring$ &\vline& $(1-0.63)/2$ & 0.1469 	 & -0.0249 	 & 2.325\\
            $\sigma$ &\vline& $(1-0.63)/4$ & -0.4098 	 & 1.829 	 & -2.87\\
            $\fcut $ &\vline& $0.3236$ 	   & -0.1331 	 & -0.2714 	 & 4.922\\
	        \hline
            \hline
        \end{tabular}
        \caption{Coefficients describing the analytical IMR waveforms 
        (see Eq.~(\ref{eq:ampParams})) in the non-spinning limit.}
        \label{tab:IMRWaveCoeffs}
    \end{center}
\end{table}

\begin{figure}[tbh]
\centering
\includegraphics[width=3.1in]{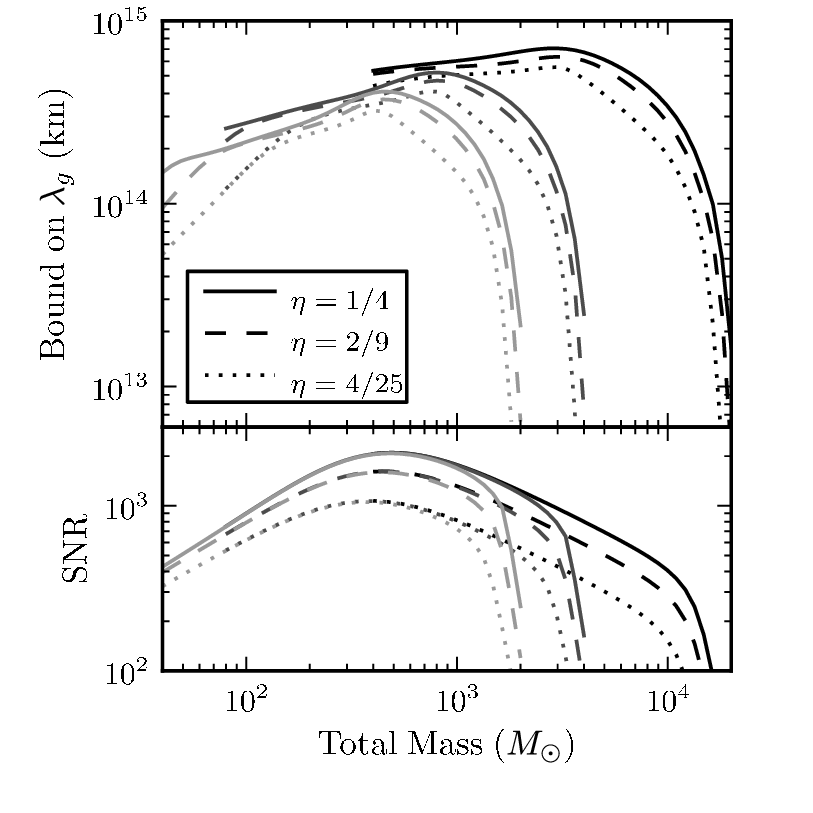}
\caption{Optimal SNR (lower panel) and the lower bound on the Compton
wavelength $\lambdag$ of graviton (upper panel) for IMR waveforms detected at
1~Gpc by the ET detector.  Black, dark grey and light grey curves correspond to
low-frequency cutoffs 1 Hz, 5 Hz, and 10 Hz, respectively, while solid, dashed,
and dotted curves correspond to mass ratios $\eta$ = 1/4, 2/9 and 4/25.}
\label{fig:ET_all}
\end{figure}

\subsection{Detector models}

Assuming that the detector noise is zero-mean stationary Gaussian, the noise 
characteristics are completely determined by its (one-sided) power spectral 
density (PSD) $S_h(f)$. 

An analytical fit to the expected noise PSD of Adv.~LIGO is given in terms of a
dimensionless frequency $x=f/f_0$~\cite{Arun:2004hn}:
\begin{equation}
S_h(f) = 10^{-49} \left [x^{-4.14} - 5 x^{-2} + 111 \Big(\frac{1 - x^2 + x^4/2}{1 + x^2/2}\Big)\right]\,,
\end{equation}
where $f_0=215$ Hz. For Adv.~LIGO, we perform our studies assuming two values
for the low-frequency cutoff, $\flow$ = 10~Hz and 20~Hz.

For the ET, the design (including the topology) is not complete. For
simplicity, we take the simplest configuration and assume ET to be a single
L-shaped interferometer with 10 km arms. The envisaged noise PSD (``ET-B
sensitivity'') of this configuration is given by the following
fit~\cite{ET:URL}: 
\begin{eqnarray}
S_h(f) & = & 10^{-50} \left [2.39\times10^{-27}\,x^{-15.64} + 0.349\,x^{-2.145} \right. \nonumber \\
& + & \left. 1.76\,x^{-0.12} + 0.409\,x^{1.1}\right]^2 \,,
\end{eqnarray}
where $f_0 = 100$ Hz. For ET, we perform our studies assuming three different 
values for $\flow$: 1 Hz, 5 Hz and 10 Hz. 

For LISA, we use the total effective non-sky-averaged spectral density,
neglecting the signal modulation due to the orbital motion of the
detector~\cite{BBW2004},
\begin{eqnarray}
S_h(f) &=& {\rm min} \left\{ S_h^{\rm NSA}(f)/{\rm exp}\left( -\kappa T^{-1} \dN/\df \right), \right. \nonumber \\
&& \left. S_h^{\rm NSA}(f) + S_h^{\rm gal}(f) \right\} + S_h^{\rm ex-gal}(f) \,,
\end{eqnarray}
where $\dN/\df = 2\times10^{-13} x^{11/3} \,{\rm Hz}^{-1}$ is the number
density of galactic white-dwarf binaries, $\kappa \simeq 4.5$ is the average
number of frequency bins that are lost when each galactic binary is fitted out,
and $T$ is the observation time. The instrumental contributions are given by,
\begin{eqnarray}
S_h^{\rm NSA}(f) &=& \left[ 9.18 \times 10^{-52} x^{-4} + 1.59 \times 10^{-41} \right. \nonumber \\
& + & \left. 9.18 \times 10^{-38} x^2 \right] \,,
\end{eqnarray}
where $f_0 = 1$ Hz. The galactic/extra-galactic white-dwarf confusion-noise 
contributions are given by,
\begin{eqnarray}
S_h^{\rm gal}(f) = 2.1 \times 10^{-45} x^{-7/3} \,, \nonumber \\
S_h^{\rm ex-gal}(f) = 4.2 \times 10^{-47} x^{-7/3} \,.
\end{eqnarray}
We start the integration at $\flow = 10^{-4}$~Hz. Binaries that we consider in
this paper would spend less than $T \simeq 5/ 256 \eta M^{5/3} (\pi
\flow)^{8/3} \simeq 8$ months in the detection band.

\begin{figure}[tbh]
\centering
\includegraphics[width=3.1in]{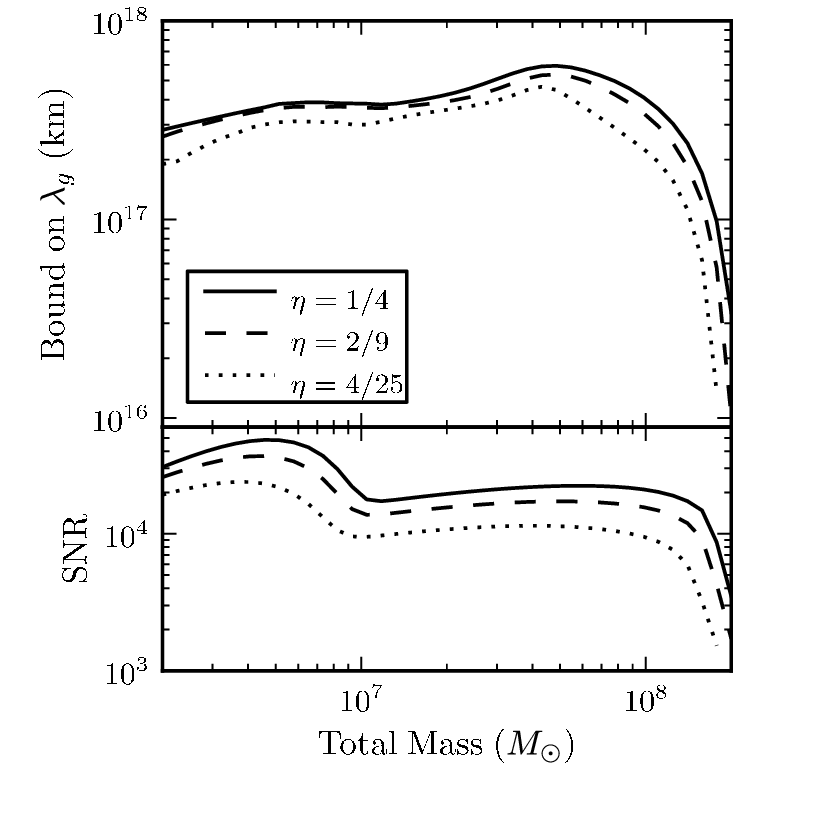}
\caption{Optimal SNR (lower panel) and the lower bound on the Compton
wavelength $\lambdag$ of graviton (upper panel) for IMR waveforms detected at
3~Gpc with the LISA detector.  The solid, dashed, and dotted curves correspond
to mass ratios $\eta$ = 1/4, 2/9, and 4/25.}
\label{fig:LISA_all}
\end{figure}

\subsection{Computing error bounds using Fisher information matrix}
\label{sec:ErrBoundsFromFishMat}

Searches for GWs from coalescing compact binaries make use of the optimal
technique of \emph{matched filtering}, which involves cross-correlating the
data $d$ with a number of theoretical templates $h(\theta^a)$ of the signal
waveforms:
\begin{equation}
\rho \equiv (d|\hat{h}) \equiv 
4\, \Re \int_{f_{\rm low}}^\infty \frac{\tilde{d}(f) \, \tilde{\hat{h}}^*(f) \, \df}{S_h(f)},
\label{eq:overlap}
\end{equation}
where $\rho$ is the SNR, tildes denote Fourier transforms, $S_h(f)$ is the
one-sided PSD of the detector noise, $f_{\rm low}$ the low-frequency cutoff of
the sensitivity, and $\hat{h} \equiv h/(h|h)^{1/2}$.  

Owing to the intrinsic randomness of the detector noise, the parameters
$\theta^a$ estimated from the search (parameters of the template giving the
maximum SNR) will be, in general, different from the ``actual'' parameters
$\vartheta^a$ of the source.  If the noise is stationary Gaussian and if the
templates are faithful representations of the true signals, then the errors
$\delta \theta^a\equiv \theta^a- \vartheta^a$ will be distributed according to
a multivariate Gaussian distribution with zero mean, whose spread can be
quantified by the elements of the variance-covariance matrix $\Sigma^{ab}$.
Specifically, the rms error in estimating the parameter $\theta^{a}$ is given
by $ \Delta \theta^{a} = \sqrt{\Sigma^{aa}}$. A relation between $\Sigma^{ab}$
and the signal is available through the Cram\`{e}r-Rao
inequality~\cite{Cramer46,Rao45}, which states that
\begin{equation}
{\bm \Sigma} \geq {\bm {\Gamma}} ^{-1},
\end{equation}
where ${\bm \Gamma}$ is the Fisher Information matrix, given by
\begin{equation}
\Gamma_{ab} \equiv \left( \frac{\partial h}{\partial \theta^{a}} \Big| 
\frac{\partial h}{\partial \theta^{b}} \right) \;, 
\end{equation}
$h$ being the signal waveform, where the inner product is defined in
Eq.(\ref{eq:overlap}).  Thus, a lower bound on the expected errors is given by:
$ \Delta \theta^{a} = \sqrt{\left(\Gamma^{-1}\right)_{aa}}$.

The parameters used in the Fisher-matrix calculation are $\theta^a \equiv \{
\ln C$, $\varphi_0$, $t_0$, $\ln M$, $\ln \eta, \beta \}$. The matrix elements
are computed by analytically computing the derivatives and numerically
evaluating the inner products. The Fisher matrix is then numerically inverted
to yield the variance-covariance matrix.  The resultant Fisher matrix is
well-conditioned throughout the parameter space we consider except at the
highest total masses, where less and less of the signal is in the detectors'
sensitive bands.

\subsection{Results and discussion}

An executive summary of results is presented in Fig.~\ref{fig:summaryplot} and
Section~\ref{sec:summary}. For ground-based detectors, the binary is assumed to
be optimally located and optimally oriented at 1~Gpc, and for LISA, the binary
is located at 3~Gpc.  For the ground-based detectors, we recompute bounds on
the mass of the graviton for various combinations of mass ratio and
low-frequency cutoff (see Figs.~\ref{fig:AdvLIGO_all} and \ref{fig:ET_all}).
For the LISA detector, we duplicate the calculations for various mass-ratio
combinations with a single low-frequency cutoff (see Fig.~\ref{fig:LISA_all}).

For the ground-based detectors, one can easily see the effect of different
low-frequency cutoffs by looking at the SNR panels of
Figs.~\ref{fig:AdvLIGO_all} and \ref{fig:ET_all}. These figures show that at
the low-mass end, the mass ratio has a stronger effect on the SNR than the
low-frequency cutoff.  This makes intuitive sense since most of the power of
the signal lies above the highest low-frequency cutoff in that regime, thus the
variation in low-frequency cutoffs doesn't significantly affect the SNR. As the
signal moves toward higher masses, the bandwidth of the waveform moves out of
the detectors' sensitive bands and the SNRs rapidly decrease.

It should be noted that the mass range giving larger bounds on $\lambda_g$ need
not correspond to the mass range giving the best SNR.  This bound is sensitive
to more than just the power of the signal present in the detectors.  Additional
degeneracy-breaking information is present at lower frequencies for a given
mass signal.  Using a smaller lower-frequency cutoff includes this information
and thus improves the bound we could place on the mass of the graviton.

The best bounds for each combination of mass ratio and low-frequency cutoff are
summarized in Table~\ref{tab:boundsSummary}. The best bound that can be placed
with Adv.~LIGO noise spectrum with low-frequency cutoff, $\flow$ = 10 Hz will
be $\sim$ 25\% better than the same obtained with $\flow$ = 20 Hz.  Similarly,
for ET, the best bounds using a configuration with $\flow$ = 1 Hz is $\sim$
70\% better than the same obtained with $\flow$ = 10 Hz.  We hope that this
information can contribute to weighing the scientific case of different
configurations of advanced detectors.

\begin{table}[tbh]
    \begin{center}
        \begin{tabular}{ccccc}
            \hline
            \hline
            $\flow$ &\vline& $\eta = 1/4$       & $\eta = 2/9$     & $\eta = 4/25$ \\
            \hline
            \multicolumn{5}{c}{Adv.~LIGO} \\
            \hline
            10 Hz  &\vline& 7.8 (360) & 	7.0 (400) & 	6.2 (360)  \\
            20 Hz  &\vline& 6.2 (220) & 	5.6 (220) & 	4.9 (200)  \\
            \hline
            \multicolumn{5}{c}{ET} \\
            \hline
            1 Hz  &\vline& 71 (3000) &  64 (3000) & 	56 (3000) \\
            5 Hz  &\vline& 52 (800)  &   47 (800) & 	41 (730)  \\
            10 Hz &\vline& 41 (440)  &   37 (440) & 	32 (400)  \\
            \hline
            \multicolumn{5}{c}{LISA} \\
            \hline
            $10^{-4}$ Hz &\vline&  5.9 (4.8 $\times 10^7$) & 	5.4 (4.8 $\times 10^7$) & 	4.7 (4.3 $\times 10^7$) \\
            \hline
            \hline
        \end{tabular}
        \caption{Best bounds on $\lambdag$ (in units of $10^{13}$ km for the case 
        of Adv.~LIGO and ET, and $10^{17}$ km for LISA) obtained from various
        combinations of mass ratio $\eta$ and low-frequency cutoff $\flow$.
        Total mass values (in $M_\odot$) giving the best bound are shown in brackets.}
        \label{tab:boundsSummary}
    \end{center}
\end{table}

We emphasize that currently there is no observational evidence for stellar-mass
black holes having mass greater than $\sim 30\, M_\odot$. Although a recently
discovered ultra-luminous X-ray source \cite{2009Natur.460...73F} represents a
possible detection of a black hole with mass $\sim 500\, M_\odot$,
observational evidence for intermediate-mass ($\sim 100 - 10^4\, M_\odot$)
black holes is still under debate \cite{MillerColbert:2004, miller-2008}.
However, population synthesis studies have suggested a number of channels
through which black-hole binaries with total mass $\sim 30 - 10^4 M_\odot$
could form~\cite{2010ApJ...715L.138B,Sadowski:2008,
Belczynski:2009xy,Brown:2007, Mandel:2007hi, imbhlisa-2006,Amaro:2006imbh}.
Although the coalescence rates of these binaries are highly uncertain, advanced
detectors \emph{might} be able to detect several of such events per
year~\cite{2010arXiv1003.2480L}.
 
Another concern is the detectability of signals deformed by the propagation
effect of the massive graviton by GR-based templates. In order to address this
question (indeed partially), we have computed the \emph{fitting factor}
(fraction of optimal SNR recovered by a suboptimal template
family)~\cite{Apostolatos:1995pj} of the GR-based IMR templates with signals
from equal-mass binaries (in the mass range $20-2000 M_\odot$, located at 1Gpc)
in Adv.~LIGO, assuming different values for the graviton mass. If we assume the
constraints on the graviton mass given by solar system tests ($\lambdag = 2.8
\times 10^{12}$ km), the fitting factor is unacceptably low (0.5 -- 0.93).
However, if the mass of the graviton is a factor of three smaller than the
solar system limit, then the fitting factors are greater than 0.97. This
dramatic change is due to the fact that the deformation of the observed signal
is proportional to the square of the Compton wavelength (see
Eq.~\ref{eq:phaseDistortion}).

\subsection{Limitations of this work}
\label{sec:limitations}

Due to the lack of a complete massive graviton theory, we have considered only
the propagation effects of a massive graviton, and have assumed that the wave
generation is correctly given by GR. If BHs in the true massive-graviton theory
are significantly different from the BHs in GR, this could affect our
estimates. Additionally, we have considered only the leading-harmonic
gravitational waveforms (in GR) produced by non-spinning BH binaries.  For the
case of binaries with high mass ratios, a considerable fraction of the emitted
energy is deposited in the higher harmonics (see, e.g., \cite{Berti:2007fi}).
Also, the current understanding is that most of the BHs in nature may be
spinning, with possibly high spin magnitudes (see, e.g.,
\cite{Volonteri:2004cf, Gammie:2003qi,Shapiro05}).  This imperfect description
of the model waveforms can introduce systematic errors in the estimated
parameters (see, e.g., ~\cite{Ajith:ParamEstimSyst}), including in the massive
graviton bounds, which this paper does not try to address.   

In this paper, the error bounds are estimated by means of the Fisher
information matrix approach. This approach has a number of limitations (see
e.g., \cite{Vallisneri:2007ev} and \cite{Zanolin:2009} for a discussion),
including the fact that this is valid only in the limit of large SNRs, does not
include various priors in the parameters (such as $\eta \in (0, 0.25]$), nor
the known issues involved in real data analysis pipelines. These limitations
are better addressed by techniques like Markov-Chain Monte-Carlo methods.

\acknowledgments 
The authors would like to acknowledge the support of the LIGO Lab, NSF grants 
PHY-0653653 and PHY-0601459, and the David and Barbara Groce Fund at Caltech.
LIGO was constructed by the California Institute of Technology and 
Massachusetts Institute of Technology with funding from the National 
Science Foundation and operates under cooperative agreement PHY-0757058. 
The authors also thank Alan Weinstein, Alessandra Buonanno and K. G. Arun 
for useful comments on the manuscript. This paper has LIGO Document Number
LIGO-P1000022-v4.

\bibliography{MassiveGraviton}

\end{document}